\documentclass[notitlepage,nofootinbib,preprintnumbers,amssymb,prl,superscriptaddress,twocolumn,nolongbibliography,reprint]{revtex4-2}
\usepackage{amsfonts,amssymb,amsmath,mathtools,graphicx,color,bm,orcidlink}
\definecolor{ultramarine}{rgb}{0.07, 0.04, 0.56}
\definecolor{cadmiumgreen}{rgb}{0.0, 0.42, 0.24}
\definecolor{indigo(dye)}{rgb}{0.0, 0.25, 0.42}
\usepackage{hyperref}
\hypersetup{
colorlinks=true,
citecolor=ultramarine,
linkcolor=cadmiumgreen,
urlcolor=indigo(dye),
}

\let\originalleft\left
\let\originalright\right
\renewcommand{\left}{\mathopen{}\mathclose\bgroup\originalleft}
\renewcommand{\right}{\aftergroup\egroup\originalright}
\usepackage{autobreak}

\allowdisplaybreaks

\begin{document}

\preprint{RESCEU-33/26}

\title{Environmental resonances in extreme-mass-ratio inspirals within boson clouds}

\author{Takuya Takahashi\,\orcidlink{0000-0002-4894-6108}}
\email{takuya.takahashi@resceu.s.u-tokyo.ac.jp}
\affiliation{Research Center for the Early Universe (RESCEU), Graduate School of Science, The University of Tokyo, Tokyo 113-0033, Japan}

\begin{abstract}
Extreme-mass-ratio inspirals (EMRIs) offer a promising probe of matter environments around massive black holes through gravitational-wave (GW) observations. We investigate orbital resonances induced by conservative gravitational perturbations from such environments. Combining a multiparameter expansion with the action-angle formalism, we derive a general resonance condition for generic bound Kerr orbits. Applying this framework to ultralight scalar clouds formed through black hole superradiance, we show that orbital resonances can arise from a single-mode cloud without requiring transitions between cloud modes. Our analysis demonstrates that these resonances can lead to significant GW dephasing, providing a new avenue for probing boson clouds through GWs and highlighting the importance of including environmental resonances in accurate EMRI waveform modeling.
\end{abstract}

\maketitle

\noindent {\bf \em Introduction---}
Extreme-mass-ratio inspirals (EMRIs) consist of a stellar-mass compact object slowly spiraling into a massive black hole (BH) through gravitational-wave (GW) emission.
These systems are key targets for space-based GW detectors~\cite{LISA:2017pwj,Hu:2017mde,TianQin:2015yph,Kawamura:2011zz}, which aim to track their phase evolution with high precision.
Beyond measuring the properties of the central BH, such observations may test gravity in the strong-field regime and reveal the influence of surrounding matter through its imprint on the orbital evolution~\cite{Barack:2006pq,Gair:2012nm,Barausse:2014pra,Cole:2022yzw,LISA:2022yao}.

Environmental effects on EMRIs have been investigated in a variety of settings, including accretion disks~\cite{Barausse:2007dy,Yunes:2011ws,Speri:2022upm}, interactions with surrounding stars~\cite{Amaro-Seoane:2011rdr,Yang:2017aht,Bonga:2019ycj}, dark matter spikes~\cite{Eda:2013gg,Kavanagh:2020cfn,Speeney:2022ryg,Rahman:2025mip}, and ultralight boson clouds~\cite{Ferreira:2017pth,Duque:2023seg,Brito:2023pyl,Takahashi:2023flk,Dyson:2025dlj,Li:2026foy}.
Considerable attention has been devoted to dissipative effects, such as dynamical friction, and their impact on the inspiral.
However, conservative perturbations can also induce significant changes in orbital evolution through resonances, as demonstrated by self-force resonances~\cite{Flanagan:2010cd} and tidal resonances driven by nearby objects~\cite{Bonga:2019ycj}.
Accounting for these effects can therefore be essential for achieving the phase accuracy required for EMRI waveform modeling~\cite{Speri:2021psr,Gupta:2021cno,Gupta:2022fbe}.
In this Letter, we combine a multiparameter expansion~\cite{Li:2022pcy,Pan:2023wau} in the mass ratio and environmental strength with the action-angle formalism~\cite{Fujita:2016igj,Gupta:2022jdt,Pan:2023wau} to describe orbital dynamics in weak matter environments and derive a general resonance condition for individual harmonic components of the matter source.

In particular, we apply this framework to ultralight scalar clouds formed through BH superradiance~\cite{Arvanitaki:2009fg,Arvanitaki:2010sy,Zouros:1979iw,Detweiler:1980uk,Dolan:2007mj,Brito:2015oca}.
Previous studies of EMRIs interacting with such clouds have examined dissipative processes associated with transitions to unbound states~\cite{Takahashi:2021yhy,Baumann:2021fkf}, as well as resonant transitions between bound states~\cite{Baumann:2018vus,Baumann:2019ztm,Tong:2022bbl,Takahashi:2023flk}.
Here, we identify orbital resonances driven by the conservative gravitational perturbation sourced by a single-mode cloud, distinct from resonances associated with cloud-mode transitions, and investigate their effects on the inspiral evolution.
The breaking of integrability and the resulting resonant dynamics have recently been studied in Ref.~\cite{Xu:2026qrg}.
We also derive the resonance condition for transitions between cloud modes along generic bound Kerr geodesics, expressed in terms of the three fundamental orbital frequencies.
Throughout this Letter, we use units in which $c=G=\hbar=1$.

\noindent {\bf \em General formalism---}
The orbit of an EMRI deviates from geodesic motion around a BH due to the gravitational self-force and is subject to an additional perturbation in the presence of a matter environment. 
Here, we adopt a multiparameter expansion to provide a general description of orbital motion in a matter environment:
\begin{align}\label{eq:metric}
    \tilde{g}_{\alpha\beta}=g_{\alpha\beta}+\xi h_{\alpha\beta}^{(1,0)}+\varepsilon h_{\alpha\beta}^{(0,1)}+{\cal O}(\xi^2,\varepsilon^2,\xi\varepsilon)~,
\end{align}
where $g_{\alpha\beta}$ is the Kerr metric in Boyer-Lindquist coordinates $x^\alpha=(t,r,\theta,\varphi)$, $\xi$ is a small parameter characterizing the strength of the matter environment, and $\varepsilon=m_p/M$ is the EMRI mass ratio, with $m_p$ and $M$ denoting the masses of the secondary object and the central BH, respectively.
Throughout this Letter, the indices $(i,j)$ denote the order $\xi^i\varepsilon^j$.

The two-timescale approach~\cite{Hinderer:2008dm} provides a natural framework for describing the orbital evolution of EMRIs by exploiting the separation between rapid orbital oscillations and slow secular evolution driven by radiation reaction.
In this framework, the integrability of geodesic motion in Kerr spacetime makes the action-angle formalism particularly well suited for analyzing orbital dynamics.
The evolution of the action-angle variables $(q^\alpha,J_\alpha)$ is governed by Hamilton's equations:
\begin{align}
    \frac{dq^\alpha}{d\tau}=&\ \Omega^\alpha(J)+\xi\Omega^\alpha_{(1,0)}(q,J;\vartheta,I)+\varepsilon\Omega^\alpha_{(0,1)}(q,J) \notag \\
    &+{\cal O}(\xi^2,\varepsilon^2,\xi\varepsilon)~, \\
    \frac{dJ_\alpha}{d\tau}=&\ \xi G_\alpha^{(1,0)}(q,J;\vartheta,I)+\varepsilon G_\alpha^{(0,1)}(q,J) \notag \\
    &+{\cal O}(\xi^2,\varepsilon^2,\xi\varepsilon)~, \label{eq:eqJ}
\end{align}
with
\begin{align}
    \Omega^\alpha_{(i,j)}=\frac{\partial H_{(i,j)}}{\partial J_\alpha}~, \quad G_\alpha^{(i,j)}=-\frac{\partial H_{(i,j)}}{\partial q^\alpha}~,
\end{align}
where the perturbed Hamiltonian at order $(i,j)$ is given by $H_{(i,j)}=-\frac{m_p}{2}h^{\alpha\beta}_{(i,j)}(x_p)u_\alpha u_\beta$, with $(i,j)=(1,0)$ or $(0,1)$.
Here, $\tau$ denotes the proper time, while $x_p^\alpha$ and $u^\alpha$ denote the position and the four-velocity of the secondary object, respectively.

The variables $(\vartheta,I)$ denote the action-angle variables of the matter constituting the environment.
A complete description of the system's evolution would require treating these variables simultaneously.
However, as we show below, the leading-order resonant contribution to the GW phase can be evaluated without explicitly evolving the matter degrees of freedom, provided that the driving harmonic remains approximately constant during resonance passage and resonant feedback is negligible.

\noindent {\bf \em Environmental resonance---}
We now consider the orbit-averaged dynamics, focusing on the contribution from $G_\alpha^{(1,0)}$ in Eq.~\eqref{eq:eqJ}.
Here, we assume that the matter stress-energy tensor takes the form $T_{\alpha\beta}^{\rm m}=e^{-i(\omega_s t-m_s\varphi)}\tilde{T}_{\alpha\beta}(r,\theta)+{\rm c.c.}$, which sources a metric perturbation of the same form, $h_{\alpha\beta}^{(1,0)}=e^{-i(\omega_s t-m_s\varphi)}\tilde{h}_{\alpha\beta}(r,\theta)+{\rm c.c.}$
For bound Kerr geodesics, the spacetime coordinates are related to the proper-time action-angle variables by~\cite{Fujita:2009bp}
\begin{align}
    t_p=q^t+\Delta t(q^r,q^\theta)~, \quad \varphi_p=q^\varphi+\Delta\varphi(q^r,q^\theta)~,
\end{align}
where $\Delta t$ and $\Delta \varphi$ are periodic functions of $q^r$ and $q^\theta$.
Substituting these relations into the perturbed Hamiltonian and Fourier expanding the remaining periodic dependence on $q^r$ and $q^\theta$, we obtain
\begin{align}
    H_{(1,0)}=\sum_{n,k}H_{nk}(J;I)e^{i q_{nk}}+{\rm c.c.}~,
\end{align}
where $q_{nk}=-\omega_s q^t+nq^r+kq^\theta+m_s q^\varphi\equiv \ell_\alpha q^\alpha$ with $\ell_\alpha=(-\omega_s,n,k,m_s)$.

Hamilton's equations then give
\begin{align}\label{eq:G_Fourier}
    G_\alpha^{(1,0)}=\sum_{n,k}G_{\alpha,nk}^{(1,0)}(J;I)e^{iq_{nk}}+{\rm c.c.}~,
\end{align}
where $G_{\alpha,nk}^{(1,0)}=-i\ell_\alpha H_{nk}$.
Since $q^\alpha\approx\Omega^\alpha\tau+q_0^\alpha$, the terms in Eq.~\eqref{eq:G_Fourier} oscillate rapidly and average out over most of the orbital evolution. 
However, a term can contribute significantly when the orbital frequencies become commensurate for a given set of mode numbers, satisfying
\begin{align}\label{eq:resonance}
    n\omega_r+k\omega_\theta+m_s\omega_\varphi=\omega_s~,
\end{align}
where $\omega_i\coloneqq\Omega^i/\Omega^t$ $(i=r,\theta,\varphi)$ are fundamental orbital frequencies with respect to Boyer-Lindquist coordinate time.
This defines the resonance condition for the effects of the matter environment on the EMRI orbit.
A well-known example is the tidal perturbation induced by a third body.
If the third body is dynamical, its perturbation frequency can be expressed as $\omega_s=n_s\omega_r'+k_s\omega_\theta'+m_s\omega_\varphi'$, where $(n_s,k_s,m_s)$ are integers labeling the mode and $\omega_i'$ are orbital frequencies of the third body. The resonance condition then reduces to the familiar form~\cite{Silva:2022blb,Kakehi:2025peb}.
We also note that stationary and axisymmetric perturbations, corresponding to $(\omega_s,m_s)=(0,0)$, lead to the same resonance condition as the self-force resonances~\cite{Flanagan:2010cd}.

The main observable consequence of the resonance is a ``jump'' in constants of motion or action variables $J_\alpha$, corresponding to an abrupt change as the EMRI orbit passes through the resonance.
Expanding the phase variable $q_{nk}$ about the resonance time $\tau_\ast$ and using $\ell_\alpha\Omega^\alpha(\tau_\ast)=0$, we obtain
\begin{align}
    q_{nk}(\tau)=&\chi_{nk}+\frac{1}{2}\Gamma_{nk}(\tau-\tau_\ast)^2+\cdots~,
\end{align}
where $\chi_{nk}=q_{nk}(\tau_\ast)$ and $\Gamma_{nk}=\ell_\alpha\dot{\Omega}^\alpha$ at $\tau=\tau_\ast$.
Thus, the change in $J_\alpha$ across a particular resonance labeled by $(n,k)$ can be well approximated by~\cite{Flanagan:2010cd}
\begin{align}
    \Delta J_\alpha\approx&\ \xi\int_{-\infty}^{\infty}G_\alpha^{(1,0)}d\tau \notag  \\
    =&\ \xi\sqrt{\frac{2\pi}{|\Gamma_{nk}|}}2{\rm Re}\left[G_{\alpha,nk}^{(1,0)}e^{i\chi_{nk}+i{\rm sgn}(\Gamma_{nk})\pi/4}\right]~. \label{eq:DeltaJ}
\end{align}
Here, all quantities in the last expression of Eq.~\eqref{eq:DeltaJ} are evaluated at the resonance time.
We note that when the resonance indices admit an overall rescaling, as in tidal resonances, only the lowest-order contribution is retained.

For a weak, isolated resonance swept through by radiation reaction, the detuning rate scales as $\Gamma\sim\varepsilon$, giving a resonance duration $\Delta \tau_{\rm res}\sim\varepsilon^{-1/2}$.
A pre-existing environmental harmonic therefore induces a fractional orbital-action jump $\Delta J/J\sim\xi/\sqrt{\varepsilon}$, which produces a fractional shift of the same order in the leading orbital frequencies.
Over a remaining inspiral time $\sim\varepsilon^{-1}$, this yields an accumulated GW phase shift
\begin{align}
    \Delta\Phi_{J}^{\rm res}\sim\frac{\xi}{\varepsilon^{3/2}}~.
\end{align}
On the other hand, if the accompanying change in environmental actions satisfies $\Delta I/I\sim\sqrt{\varepsilon}$, its contribution to the orbital frequencies enters only through $\xi\Omega_{(1,0)}$, yielding an additional phase shift $\Delta\Phi_I^{\rm res}\sim\xi/\sqrt{\varepsilon}$.
A secular interaction such as gravitational dynamical friction gives $\Delta\Phi_{\rm int}\sim\xi/\varepsilon$ if its contribution to $\dot{J}/J$ scales as $\xi\varepsilon$.
Under these assumptions, provided that resonant feedback on the frequency sweep is negligible, the direct orbital resonance is parametrically dominant among these environmental contributions.

For a dynamical environment, resonant excitation of environmental degrees of freedom can invalidate the assumption of a prescribed, nearly constant driving harmonic.
The resulting orbital resonance is part of the associated exchange of energy and angular momentum and should not be counted independently.
If environmental evolution or orbital backreaction substantially modifies the driving amplitude or resonance duration, as in resonance capture or floating, the preceding hierarchy need not apply, and the environment and orbit must be solved self-consistently.

\noindent {\bf \em Scalar cloud---}
As a concrete example of the matter constituting the environment, we consider an ultralight scalar field, whose stress-energy tensor is given by
\begin{align}
    T_{\alpha\beta}=\nabla_\alpha\phi\nabla_\beta\phi+g_{\alpha\beta}\left(-\frac{1}{2}\nabla_\mu\phi\nabla^\mu\phi-\frac{\mu^2}{2}\phi^2\right)~,
\end{align}
where $\mu$ is the mass of the scalar field.
The field equation admits a separable solution of the form~\cite{Brill:1972xj}
\begin{align}
    \phi=e^{-i(\omega t-m\varphi)}R_{lm\omega}(r)S_{lm\omega}(\theta)+{\rm c.c.}
\end{align}
Here, we consider quasi-bound states satisfying purely ingoing boundary conditions at the horizon and exponentially decaying boundary conditions at spatial infinity.
The eigenfrequencies are generally complex, and the modes become unstable when the superradiant condition $0<\omega<m\Omega_H$ is satisfied~\cite{Dolan:2007mj}, where $\Omega_H=a/(2Mr_+)$, $r_+=M+\sqrt{M^2-a^2}$, and $a$ is the BH spin parameter.
We now investigate the resonant effects on the EMRI orbit induced by a scalar cloud formed through this superradiant instability.
In the following, we neglect the imaginary part of $\omega$ for simplicity, since the cloud growth time is much longer than the inspiral timescale.
For the scalar cloud, we choose $\xi=M_c/M$, where $M_c$ denotes the cloud mass.

We first consider a scalar cloud consisting of a single mode and focus on its contribution to the metric perturbation.
The stress-energy tensor sourcing this perturbation is quadratic in $\phi$ and can therefore be decomposed into two parts as
\begin{align}
    T_{\alpha\beta}=&\tilde{T}^{(0\omega,0m)}_{\alpha\beta} (r,\theta) \notag \\
    &+\left[e^{-2i(\omega t-m\varphi)}\tilde{T}_{\alpha\beta}^{(2\omega,2m)}(r,\theta)+{\rm c.c.}\right]~.
\end{align}
Thus, there are two distinct sources corresponding to $(\omega_s,m_s)=(0,0)$ and $(\omega_s,m_s)=(2\omega,2m)$, each of which can give rise to a resonance.
The former leads to the same resonance condition as that for self-force resonances, while the oscillatory component of the scalar cloud gives rise to additional resonances.

\begin{figure}[t]
\centering
\includegraphics[width=0.8\linewidth]{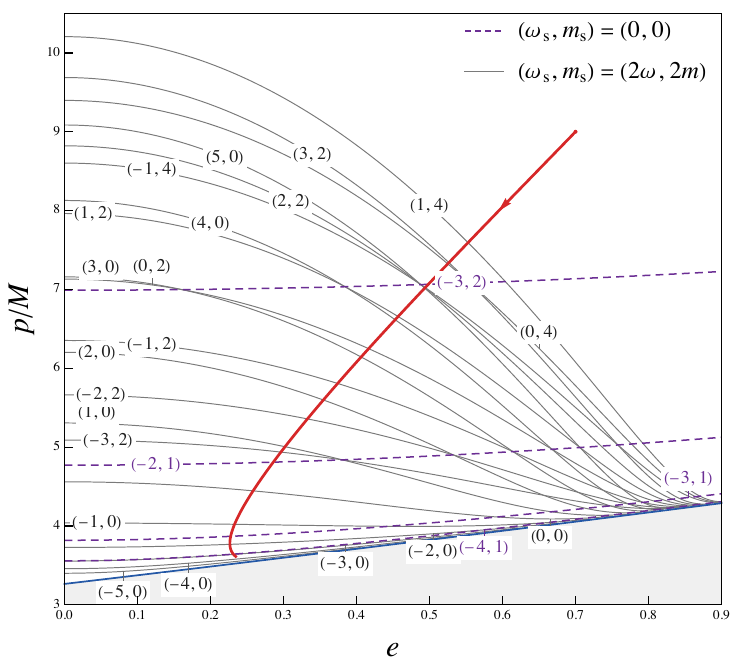}
\caption{Resonance contours induced by a scalar-cloud environment, shown in the $e$--$p$ plane at fixed inclination $\iota=50^\circ$, with $M\mu=0.1$ and $a/M=0.9$. Purple dashed and gray solid curves correspond to $(\omega_s,m_s)=(0,0)$ and $(2\omega,2m)$, respectively, and are labeled by $(n,k)$. The red curve shows the projection of an adiabatically evolving EMRI trajectory onto this plane, while the blue curve marks the separatrix.}
\label{fig:resonance_contours}
\end{figure}

In Fig.~\ref{fig:resonance_contours}, we show the resonance contours associated with the two components in the $e$--$p$ plane, based on Eq.~\eqref{eq:resonance}.
Here, $p$ is the semi-latus rectum, $e$ is the orbital eccentricity, and the orbital inclination $\iota$ is fixed at $50^\circ$.
For the scalar field, we consider the fastest-growing mode with $l=m=1$ and set $M\mu=0.1$ and $a/M=0.9$.
The eigenfrequency is computed using Leaver's method~\cite{Dolan:2007mj,Leaver:1986vnb}, while the orbital frequencies of the EMRI are calculated using the Black Hole Perturbation Toolkit~\cite{BHPToolkit}.
We plot the resonance contours up to $|n|+|k|\leq5$.
The equatorial reflection symmetry of the source implies that $H_{nk}=0$ for odd $k$, so only even-$k$ resonance contours are shown for the $(2\omega,2m)$ sector. 
For the $(0,0)$ sector, contours labeled by odd $k$ are retained because their even integer multiples correspond to the same resonance surface.

In a binary system with a scalar cloud, tidal interactions are known to induce transitions between the energy levels of the cloud~\cite{Baumann:2019ztm}.
In the present formalism, this corresponds to a stress-energy tensor constructed from multiple modes, including those that are not initially excited.
Labeling the two relevant modes by 1 and 2, the corresponding perturbation is sourced by the cross terms in the stress-energy tensor, and the resonance condition for level transitions along generic orbits is given by
\begin{align}
    n\omega_r+k\omega_\theta+\Delta m\omega_\varphi=\Delta\omega~,
\end{align}
where $\Delta \omega=\omega_1-\omega_2$ and $\Delta m=m_1-m_2$.
This condition generalizes the Schwarzschild result of Refs.~\cite{Xu:2026cky,Xu:2026aic}.
As extensively studied in the literature, such resonances can lead to phenomena such as floating or shrinking orbits.
We expect that these resonances often cannot be treated as weak perturbations and therefore require a self-consistent treatment of the orbital evolution and scalar cloud dynamics.
We leave a detailed analysis for future work.

\noindent {\bf \em Impact on the orbit---}
Focusing on resonances arising from the $(\omega_s,m_s)=(0,0)$ sector of a single-mode cloud, we examine their effects on the orbital evolution.
We evolve the EMRI orbit adiabatically under gravitational radiation reaction according to~\cite{Ganz:2007rf,Isoyama:2018sib}
\begin{align}
    \left<\frac{dC^A}{dt}\right>=-{\cal F}^A_{\rm RR}(C)~, \quad \left<\frac{dz^a}{dt}\right>=(T^{-1})^a_{\ A}\left<\frac{dC^A}{dt}\right>~,
\end{align}
where $C^A=(E,L_z,Q)$ are the orbital energy, axial angular momentum, and Carter constant, respectively, and $z^a=(p,e,\iota)$ with $T^A_{\ a}=\partial C^A/\partial z^a$.
The fluxes ${\cal F}^A_{\rm RR}$ are computed using the Kerr Teukolsky formalism implemented in \texttt{pybhpt}~\cite{Nasipak_pybhpt_2026,Nasipak:2025tby}.
An example of the resulting inspiral trajectory is shown by the red curve in Fig.~\ref{fig:resonance_contours}.

For the $(0,0)$ sector, the energy and the angular momentum are conserved, whereas the Carter constant can vary.
Its instantaneous rate of change is given by~\cite{Yang:2017aht}
\begin{align}\label{eq:dQdtau}
    \frac{1}{\xi}\frac{dQ}{d\tau}=2u_\theta a_\theta-2a^2\cos^2\theta u_t a_t+2\cot^2\theta u_\varphi a_\varphi~,
\end{align}
where $a^\alpha=-\frac{1}{2}(g^{\alpha\beta}+u^\alpha u^\beta)(2h^{(1,0)}_{\beta\lambda;\rho}-h^{(1,0)}_{\lambda\rho;\beta})u^\lambda u^\rho$ is the acceleration induced by the environment.
From Eq.~\eqref{eq:G_Fourier}, the orbit-averaged environmental forcing can be written as
\begin{align}
    \left<G_\alpha^{(1,0)}\right>\approx G_{\alpha,nk}^{(1,0)}(J;I)e^{i(nq_0^r+kq_0^\theta)}+{\rm c.c.}
\end{align}
Since the Carter constant is a function of the orbital actions, its averaged rate of change $\left<dQ/dt\right>$ inherits the dependence on the initial phase combination.

\begin{figure}[t]
\centering
\includegraphics[width=1.0\linewidth]{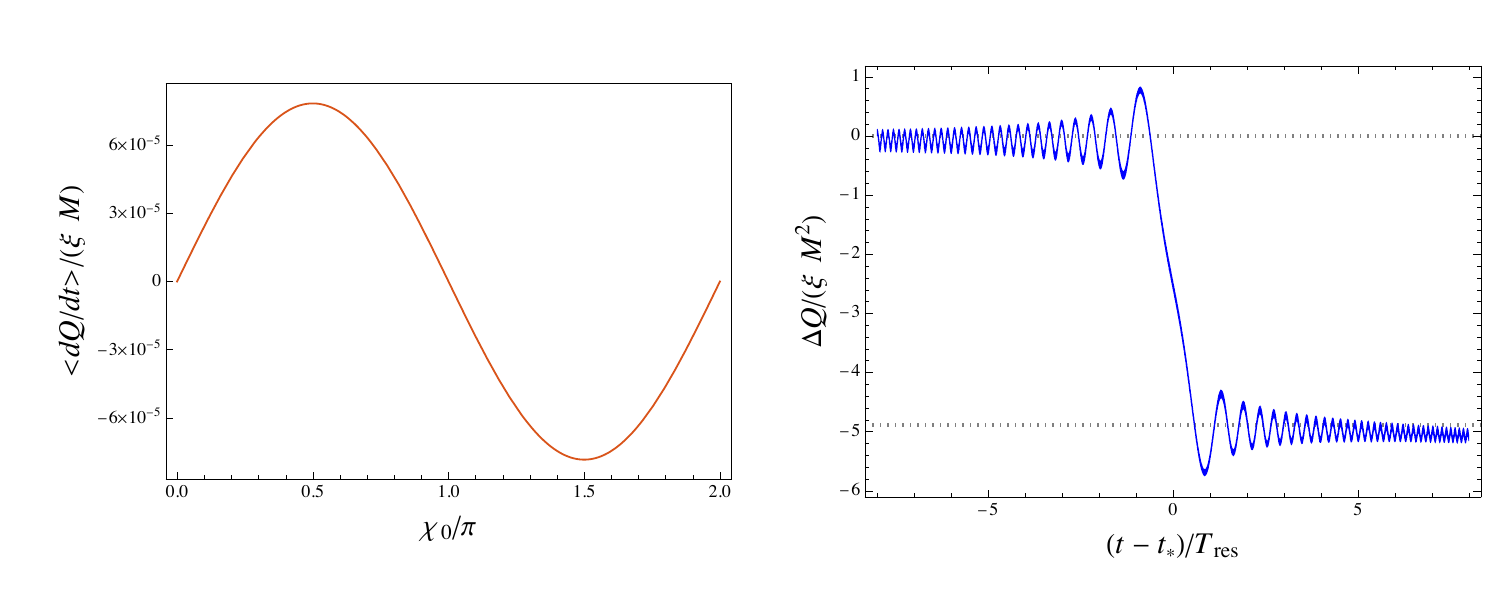}
\caption{Cloud-induced changes in the Carter constant at the $(n,k)=(3,-2)$ resonance for $a/M=0.9$, $M\mu=0.2$, $\xi=M_c/M=10^{-4}$, and $\varepsilon=10^{-5}$. \textbf{Left:} Initial phase dependence of the orbit-averaged rate $\left<dQ/dt\right>$. \textbf{Right:} Difference $\Delta Q(t)$ between orbital evolutions with and without the cloud, with the initial phase chosen to maximize the jump amplitude. Here $T_{\rm res}=\sqrt{2\pi/|d^2\chi/dt^2|_{t=t_\ast}}$, with $\chi=3q^r-2q^\theta$. The dotted lines indicate zero and the analytic jump prediction.}
\label{fig:cloud00_resonance}
\end{figure}

For $h_{\alpha\beta}^{(1,0)}$, we adopt the post-Newtonian approximation of Ref.~\cite{Xu:2026qrg}, identifying its harmonic coordinates with those of the Kerr background~\cite{Sopuerta:2011te}.
Further details are provided in the Supplemental Material.
Fig.~\ref{fig:cloud00_resonance} illustrates the effect of a scalar cloud with $M\mu=0.2$ and $\xi=10^{-4}$ on the Carter constant at the $(n,k)=(3,-2)$ resonance.
We set $a/M=0.9$ and consider an orbit with $e\simeq0.50$, $\iota\simeq50^\circ$, and $p/M\simeq7.10$ at resonance.
The left panel shows the dependence of the orbit-averaged rate $\left<dQ/dt\right>$ on the initial resonant phase $\chi_0=3q_0^r-2q_0^\theta$.
The right panel shows $\Delta Q(t)=Q_{\rm RR+cloud}(t)-Q_{\rm RR}(t)$, obtained by integrating the orbital evolution equations with and without the cloud using the method of osculating Kerr geodesics~\cite{Gair:2010iv,Pound:2007th}, including radiation reaction in both cases, with $\varepsilon=10^{-5}$.
Both trajectories start from identical initial conditions, with the phase chosen to maximize the magnitude of the jump predicted on the radiation-reaction background.
The resulting jump is consistent with the prediction of the jump formula~\eqref{eq:DeltaJ}.

\begin{figure}[t]
\centering
\includegraphics[width=1.0\linewidth]{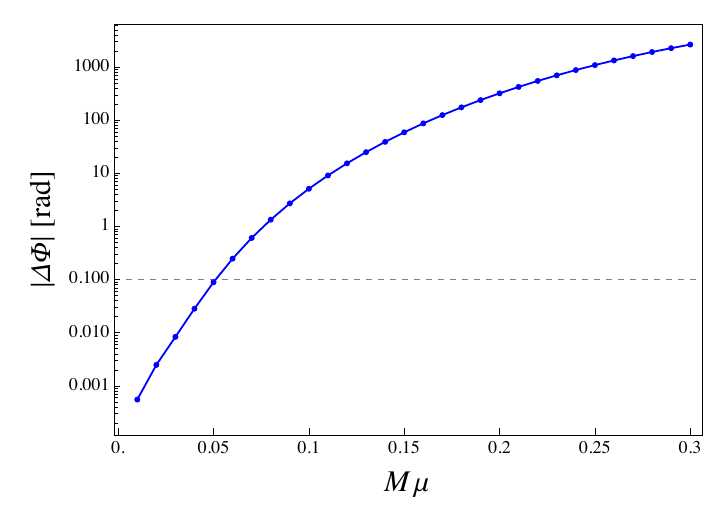}
\caption{Absolute GW dephasing $|\Delta\Phi|$ accumulated from the $(n,k)=(3,-2)$ resonance to plunge as a function of $M\mu$, assuming the maximum resonance jump. We fix $a/M=0.9$, $\xi=M_c/M=10^{-4}$, and $\varepsilon=10^{-5}$, with the same orbital configuration as in Fig.~\ref{fig:cloud00_resonance}. The gray dashed line indicates a reference dephasing of 0.1 rad.}
\label{fig:cloud00_dephasing}
\end{figure}

Finally, to estimate the impact on EMRI waveforms, we evaluate the azimuthal GW dephasing,
\begin{align}
    \Delta\Phi=\int_{t_\ast}^{T_{\rm plunge}} 2\Delta\omega_\varphi\ dt~.
\end{align}
Here $\Delta\omega_\varphi$ is the difference between the azimuthal frequencies along adiabatically evolving orbits with and without the maximum jump $\Delta Q$ from Eq.~\eqref{eq:DeltaJ}.
The validity of the weak-resonance approximation is assessed in the Supplemental Material.
This phase difference accumulates from the resonance crossing at $t=t_\ast$ until either orbit reaches a semilatus rectum $0.05M$ above the separatrix value.
Fig.~\ref{fig:cloud00_dephasing} shows the GW dephasing as a function of $M\mu$, with all other parameters and the orbital configuration identical to those in Fig.~\ref{fig:cloud00_resonance}.
The results suggest that increasing $M\mu$ leads to a more compact cloud and a stronger perturbation at the resonance location, enhancing the resonant effect on the orbit.

\noindent {\bf \em Conclusions---}
We have presented a general analysis of orbital resonances induced by conservative environmental perturbations of EMRIs using the action-angle formalism.
Applying this framework to scalar clouds, we have explicitly derived the resonance conditions and analyzed the impact of the stationary sector on orbital evolution.
For the setup considered in Fig.~\ref{fig:cloud00_dephasing}, the resulting GW dephasing can exceed 0.1 rad, a benchmark adopted in previous studies to assess potential observability with LISA at signal-to-noise ratios of about 20--30~\cite{Bonga:2019ycj,Gupta:2021cno}.
These results demonstrate the need to incorporate this class of resonances into accurate EMRI waveform models in matter environments.

Several directions remain for future work.
The oscillatory $(2\omega,2m)$ sector admits a richer set of resonances than the stationary, axisymmetric sector, and their cumulative effect warrants investigation even if individual contributions are small~\cite{Ferreira:2017pth}.
Their dependence on the cloud frequency may also help distinguish them from self-force resonances.
A fully relativistic treatment of the cloud-induced metric perturbation would also improve the accuracy of the present estimates.
Beyond the weak-resonance regime, it would be interesting to study passage through and possible capture into resonance islands in perturbed BH spacetimes~\cite{Pan:2023wau,Strateny:2025zkd,Xu:2026qrg}.
Ultimately, a complete waveform model will require a self-consistent evolution of the orbit and the cloud.
Further extensions include self-interacting axion clouds with multiple excited modes~\cite{Takahashi:2024fyq} and vector clouds~\cite{Cao:2024wby}.
Application of this framework to other environments remains to be explored.

\noindent {\bf \em Acknowledgments---}
The author thanks Takahiro Tanaka and Hidetoshi Omiya for useful discussions, Richard Brito and Qi-Xuan Xu for helpful comments on the manuscript, and Ryuichi Fujita for sharing numerical flux data for an independent check of the fluxes computed in this work.
The author acknowledges the use of Codex (OpenAI) to assist with code development.
This work makes use of the Black Hole Perturbation Toolkit~\cite{BHPToolkit}.
This work is supported by JSPS (Japan Society for the Promotion of Science) KAKENHI Grant Nos.~JP25KJ0067 and JP25K17397.

\section{Supplemental Material}
\label{suppmaterial}

\setcounter{equation}{0}
\renewcommand{\theequation}{S\arabic{equation}}
\setcounter{figure}{0}
\renewcommand{\thefigure}{S\arabic{figure}}
\setcounter{table}{0}
\renewcommand{\thetable}{S\arabic{table}}

\subsection{Treatment of the metric perturbation}
We consider the environmental perturbation $h_{\alpha\beta}^{(1,0)}$ defined in Eq.~\eqref{eq:metric}, restricting attention to the stationary, axisymmetric $(\omega_s,m_s)=(0,0)$ sector of the scalar cloud.
We retain the terms linear in $\xi$ in the first post-Newtonian metric of Ref.~\cite{Xu:2026qrg}, including the BH-cloud interaction terms.
Below, we abbreviate $h_{\alpha\beta}^{(1,0)}$ as $h_{\alpha\beta}$.

In spherical variables $y^A=(T,\rho_H,\theta_H,\varphi_H)$ associated with the harmonic Cartesian coordinates, the cloud perturbation takes the form
\begin{align}
    h_{AB}^Hdy^A dy^B=&\left(-2u-2\psi+\frac{4M}{\rho_H}u\right)dT^2 \notag \\ &-8w\sin^2\theta_H dT d\varphi_H \notag \\
    &-2u(d\rho_H^2+\rho_H^2d\theta_H^2+\rho_H^2\sin^2\theta_H d\varphi_H^2)~.
\end{align}
Here, $u$, $\psi$, and $w$ are the potentials given in Appendix B of Ref.~\cite{Xu:2026qrg}, divided by $\xi=M_c/M$.
Here, $M_c$ denote the mass parameter used to normalize the hydrogenic cloud profile, which agrees with the gravitational mass of the cloud at Newtonian order.
The scalar potentials contain monopole and quadrupole contributions, while $w$ describes the azimuthal current.

We identify these harmonic Cartesian coordinates with those of the Kerr background specified in Ref.~\cite{Sopuerta:2011te}.
Their associated spherical variables are related to Boyer-Lindquist coordinates $x^\alpha=(t,r,\theta,\varphi)$ by
\begin{equation}
\begin{aligned}
    &T=t~, \quad \rho_H=\sqrt{(r-M)^2+a^2\sin^2\theta}~,\\
    &\cos\theta_H=\frac{(r-M)\cos\theta}{\rho_H}~, \quad \varphi_H=\varphi-\gamma(r)~, \\
    &\gamma'(r)=\frac{aM^2}{(r^2-2Mr+a^2)\{(r-M)^2+a^2\}}~.
\end{aligned}
\end{equation}
The harmonic condition applies to the Cartesian coordinates, rather than to the spherical variables individually.
This identification specifies our model for embedding the post-Newtonian cloud field in the Kerr background.

We transform the perturbation according to
\begin{align}
    h_{\alpha\beta}^{\rm BL}=\frac{\partial y^A}{\partial x^\alpha}\frac{\partial y^B}{\partial x^\beta}h_{AB}^H~,
\end{align}
and consistently truncate both the potentials and the coordinate transformation at the retained post-Newtonian orders, keeping terms through ${\cal O}(v^4)$ in $h_{tt}$, ${\cal O}(v^3)$ in $h_{ti}$, and ${\cal O}(v^2)$ in $h_{ij}$.
These orders refer to Cartesian components or their leading-order orthonormal counterparts.
With $a/M={\cal O}(1)$ and $M/r={\cal O}(v^2)$, the radial shift $\rho_H=r-M+{\cal O}(rv^4)$ gives
\begin{align}
    u(\rho_H,\theta_H)=u(r,\theta)-M\partial_r u(r,\theta)+{\cal O}(v^6)~.
\end{align}
The remaining corrections from the coordinate map contribute beyond the retained component orders.
Omitting the superscript BL, the resulting nonzero components are
\begin{equation}
\begin{aligned}
    &h_{tt}=-2u-2\psi+\frac{4M}{r}u+2M\partial_r u~,\\
    &h_{t\varphi}=-4w\sin^2\theta~, \\
    &h_{rr}=-2u~, \\
    &h_{\theta\theta}=-2r^2u~, \\
    &h_{\varphi\varphi}=-2r^2\sin^2\theta u~.
\end{aligned}
\end{equation}
All potentials in these expressions are evaluated at $(r,\theta)$.
In particular, $2M\partial_r u$ is required by the coordinate transformation at the retained order.

The post-Newtonian truncation is applied only to the cloud perturbation and its coordinate transformation.
The Kerr background is retained exactly when computing the orbital motion and the environmental acceleration.

\subsection{Validity of the weak-resonance approximation}
The stationary-phase jump formula~\eqref{eq:DeltaJ} evaluates the resonant forcing along an inspiral driven by radiation reaction (RR)~\cite{Flanagan:2010cd}.
Its validity requires that the forcing only weakly modifies the passage through resonance.
For the $(n,k)=(3,-2)$ resonance, we define the detuning $\delta=3\omega_r-2\omega_\theta$, where the frequencies are measured with respect to coordinate time $t$.
The RR-driven rate of change of the detuning and the resonance timescale are
\begin{align}
    \beta=\left.\frac{d\delta}{dt}\right|_{t=t_\ast}~, \quad T_{\rm res}=\sqrt{\frac{2\pi}{|\beta|}}~,
\end{align}
where $t_\ast$ is the resonance crossing time on the RR trajectory.

For the resonance induced by the $(\omega_s,m_s)=(0,0)$ sector considered in the main text, only the Carter constant $Q$ undergoes a resonant jump at leading order, while $E$ and $L_z$ remain unchanged.
Applying the resonant-phase analysis of Ref.~\cite{vandeMeent:2013sza} and neglecting direct environmental corrections to the angle evolution, the weak-feedback condition becomes
\begin{align}
    {\cal R}\equiv\frac{|\partial_Q\delta||\Delta Q|_{\rm max}}{|\beta|T_{\rm res}}\ll1~.
\end{align}
Here, $\partial_Q\delta$ is evaluated at fixed $E$, $L_z$ at the resonance, and $|\Delta Q|_{\rm max}$ is the maximum jump amplitude predicted by the lowest harmonic stationary-phase formula~\eqref{eq:DeltaJ}.
This ratio compares the maximum resonant contribution to $d\delta/dt$ with the RR contribution.
For fixed cloud and orbital configurations, ${\cal R}\propto\xi/\varepsilon$, but its magnitude also depends on the resonant forcing and frequency gradient.

For the orbital configuration used in Fig.~\ref{fig:cloud00_dephasing}, with $a/M=0.9$, $\xi=10^{-4}$, and $\varepsilon=10^{-5}$, we obtain ${\cal R}\simeq0.025$ at $M\mu=0.2$ and ${\cal R}\simeq0.23$ at $M\mu=0.3$.
Thus, the weak-resonance condition is well satisfied for $M\mu=0.2$, while it may be marginal for $M\mu=0.3$.
In general, however, whether the condition is satisfied also depends on $\xi$ and $\varepsilon$.


\bibliographystyle{mybibstyle}
\bibliography{ref}

\end{document}